\documentstyle[preprint,aps,psfig]{revtex}   
\include{amsmath}
\begin{document}

\title{Rate of three-body recombination of hydrogen\\
molecules during primordial star formation}

\author{Robert C. Forrey}

\address{Department of Physics, Penn State University, 
Berks Campus, Reading, PA 19610-6009}

\date{\today}
\maketitle

\vspace*{-.3in}
\begin{abstract}
Astrophysical models of primordial star formation 
require rate constants for three-body recombination as input.
The current status of these rates for H$_2$ due to 
collisions with H is far from satisfactory, with
published rate constants showing orders of 
magnitude disagreement at the temperatures 
relevant for H$_2$ formation in primordial gas. 
This letter presents an independent calculation
of this recombination rate constant as a function
of temperature.  An analytic expression is
provided for the rate constant which should be more 
reliable than ones currently being used in 
astrophysical models.
\end{abstract}


\newpage

A recent study by Turk et al. \cite{turk} on the effects of varying the  
three-body recombination (TBR) rate of hydrogen during primordial star 
formation concluded that 
``the uncertainty in the three-body H$_2$ formation rate significantly
limits our ability to model the density, temperature, and velocity structure
of the gas close to the center of the collapse" and
``the uncertainty in the outcome of collapse caused by our poor knowledge 
of the three-body H$_2$ formation rate coefficient cannot be so easily
dealt with and represents a major limitation on our ability to accurately 
simulate the formation of the first stars in the universe." 

In this study,
three different published rate constants were used for TBR of H$_2$ due to 
collision with H. Two of these rate constants were based on shock tube 
measurements of Jacobs et al. \cite{jacobs} who gave analytic expressions 
for TBR and the inverse process of collision induced dissociation (CID).
The TBR rate adopted by Palla et al. \cite{palla} was identical to the
expression given by \cite{jacobs}. Flower and Harris \cite{flower} used
the CID expression given by \cite{jacobs} together with their own
determination of the equilibrium constant to derive a very different TBR rate constant.
The discrepancy between these two TBR rate constants, therefore, is due to 
the adopted equilibrium constants that were used. The equilibrium constant 
used by Flower and Harris \cite{flower}
relies on the Saha equation and their determination of the H$_2$ partition 
function. When fitted to a simple temperature dependent function,
Flower and Harris found that their value is approximately 4.5 times 
greater than a similar fit obtained from the JANAF Thermochemical Tables 
at a temperature of 1000 K. Using this equilibrium constant together with
the CID expression given in \cite{jacobs} yielded a TBR rate constant 
which was approximately 6 times larger at T=1000 K than the TBR expression 
given in \cite{jacobs}.

It appears that there are two separate reasons for the factors of 4.5 and 6
discrepancies at 1000 K. The first reason is fitting error. The experiments 
were performed in a temperature range of 2900-4700 K. When the same comparisons
for the equilibrium constant and TBR rate are made in this experimental temperature 
range, both discrepancies are reduced to a factor of 4 over the whole range. 
In order to see this, the exact H$_2$ partition function must be used to
compute the equilibrium constant, not the fitted value \cite{flower} which is valid only for $T<2000$ K.
The second reason is the choice of atomic partition function which was assumed to be 2 
for hydrogen in its 1s $^2$S ground state \cite{flower}. This choice accounts for the
nuclear spin degeneracy but not the electron spin degeneracy. Because 
the formation of H$_2$ occurs in the singlet electronic ground state, the electron spin 
degeneracy was assumed to be unity \cite{flower2}. The present author disagrees with this 
assumption because 3/4 of the atomic collisions would still occur on the repulsive triplet 
electronic state even though they do not react. 
In order to account for only the 1/4 of collisions which can react, 
the atomic partition function used by Flower and Harris \cite{flower} would need
to be increased by a factor of 2 which would reduce their equilibrium constant 
by a factor of 4. Therefore, the TBR rate of Flower and Harris \cite{flower} 
would also need to be reduced by this same factor of 4.  This would, upon use
of the exact H$_2$ partition function, bring it 
into excellent agreement with the TBR rate of Jacobs et al. \cite{jacobs} and 
Palla et al. \cite{palla} at temperatures in the experimental range.

While this resolves the discrepancy between two of the three rate constants used in
the Turk et al. \cite{turk} study, it does not imply that the rate constants are
reliable. The experimental paper \cite{jacobs} upon which these TBR rate constants
are based, cautions that their data lies in the middle of a range of other 
experimental data whose values vary over an order magnitude. Furthermore,
there is no justification for extrapolating collisional data obtained over 
a small range of temperatures, to temperatures that lie outside that range.
All three of the TBR rates considered in the Turk et al. \cite{turk} study
are based on extrapolations for temperatures between 300 and 2900 K. 
The third and smallest TBR rate,
those of Abel et al. \cite{abel}, uses an extrapolation of classical 
trajectory calculations performed by Orel \cite{orel} at temperatures
below 300 K.  This extrapolation assumes an inverse temperature dependence 
which resembles that of Palla et al. \cite{palla} and consequently
has a sudden change in slope at 300 K.
It is desirable, therefore, to perform an independent calculation
which does not rely on extrapolations and is reliable over the 
temperature range required by the astrophysical models.

In this letter, we report results of quantum mechanical calculations
of TBR rate constants for the collision of three hydrogen atoms
in the temperature range $300 <  T < 10,000$ K. 
We use a Sturmian representation which provides a quadrature 
of the two-body continuum and may be used to generate a complete set 
of states within any desired TBR pathway \cite{bob}. 
The effective TBR and CID rate constants at local thermodynamic 
equilibrium (LTE) may then be defined by 
\begin{equation} 
k_r\equiv\sum_{b,u}k_{u\rightarrow b}
\frac{g_u\exp(-E_u/k_BT)}{Q_H^2Q_T}
\label{kr}
\end{equation}
\begin{equation}
k_d\equiv\sum_{b,u}k_{b\rightarrow u}
\frac{g_b\exp(-E_b/k_BT)}{Q_{H_2}}
\label{kd}
\end{equation}
where $b$ designates a bound state with energy $E_b$ and $u$ designates
an unbound state with energy $E_u$. These states are defined by their
associated vibrational and rotational quantum numbers $v$ and $j$.
The statistical factors are given by $g=(2j+1)(2I+1)$ with the
nuclear spin $I=0$ for para-H$_2$ and $I=1$ for ortho-H$_2$.
With this definition for $g$, the atomic partition function
is $Q_H=4$ as described above. 
The molecular partition function $Q_{H_2}$ and the 
translational partition function $Q_T$ are defined by
\begin{equation}
Q_{H_2}=\sum_{b}g_b\exp(-E_b/k_BT)
\label{QH2}
\end{equation}
\begin{equation}
Q_T=\frac{(\pi m k_BT)^{3/2}}{h^3}
\label{QT}
\end{equation}
where $h$ is Planck's constant, $k_B$ is Boltzman's constant, 
$T$ is the temperature, and $m$ is the mass of H.
Detailed balance of the rate coefficients $k_{b\rightarrow u}$
and $k_{u\rightarrow b}$
may be used to show that the above definitions 
yield the statistical Saha equation 
\begin{equation}
\frac{k_r}{k_d}=\frac{[H_2]}{[H]^2}=\frac{Q_{H_2}}{Q_H^2Q_T}
\label{saha}
\end{equation}
for the thermalization of the continuum. For the results
reported here, we assume that the system is in equilibrium. 
Corrections for non-LTE conditions are estimated to be small.
The calculations use an energy sudden approximation which was tested
for He+H+H by comparing with coupled states calculations \cite{bob}. 
Results from the two methods showed good agreement at temperatures 
greater than 600 K.
The calculations also assume that the atoms are distinguishable
which should be a good approximation at the high temperatures
under consideration and is consistent with classical calculations. 
The BKMP2 \cite{bkmp2} potential energy surface (PES) was employed 
for the calculations. Previous quasiclassical calculations by Esposito
and Capitelli \cite{esposito} found that the BKMP2 PES gave very 
similar results compared to the LSTH PES \cite{lsth} over the 
same temperature range considered in the present work.
 
Figure 1 shows the results of the present calculations
together with the three TBR rate constants considered in the
study of Turk et al. \cite{turk}. The present results are 
smaller than those of Flower and Harris \cite{flower}
and Palla et al. \cite{palla} over the entire temperature
range shown. These curves are both based on the experimental 
data of Jacobs et al. \cite{jacobs} as discussed above. 
Compared to the Palla et al. curve, we find good agreement 
with the extrapolation at high temperatures but poor agreement 
at low temperatures.
We see the opposite effect when compared to the Abel et al. curve.
The present calculations produce a much flatter temperature
dependence than the other three curves.
This flat temperature dependence 
is in good agreement with the quasiclassical 
results of Esposito and Capitelli \cite{esposito} which
were also computed using the BKMP2 PES.
The magnitude of the present results agree with those
of Esposito and Capitelli
to within a factor of 2 over the entire temperature
range shown. They also agree to within a factor of 2
with Jacobs et al. \cite{jacobs} and Orel \cite{orel}.
Therefore, the factor of $\sim 100$ uncertainty which
was introduced by the various extrapolations is
estimated to be reduced to a factor of $\sim 2$
when the present results are used.

The results of the present calculations may be conveniently
expressed by the function
\begin{equation}
k_r=6\times 10^{-32}T^{-1/4}+2\times 10^{-31}T^{-1/2}
\end{equation}
using the same units as in Figure 1.  This expression      
is virtually identical to the calculations for $T=600-6000$ K
and not too different at the endpoints of the range shown in
the figure. This analytic function should be reliable
for temperatures required by the hydrodynamics simulations.
Based on previous simulations \cite{turk} it is expected that
the gravitational collapse will produce a gas distribution
which is somewhere in between the spherical distribution 
obtained using the Abel et al. \cite{abel} rate constant 
and the bar-shaped distribution obtained using the 
Palla et al. \cite{palla} rate constant.

It is noteworthy that recent numerical simulations of 
gravitational fragmentation by Clark et al. \cite{clark}
adopted a TBR rate constant which was in between that of
Palla et al. \cite{palla} and Abel et al. \cite{abel}.
This rate constant was due to Glover \cite{glover}
who used CID rate constants of Martin et al. \cite{martin}
together with the fitted equilibrium constant of 
Flower and Harris \cite{flower} which as noted above 
is 4 times too large when $T<2000$ K. 
It is recommended that future astrophysical 
simulations of gravitational collapse employ the revised equilibrium
and TBR rate constants accordingly.

This work was supported by the NSF grant No. PHY-1203228.
The author would like to thank Prof. Phillip Stancil for
a careful reading of the manuscript and acknowledge 
Prof. David Flower for helpful communications.

\newpage
\centerline{\psfig{figure=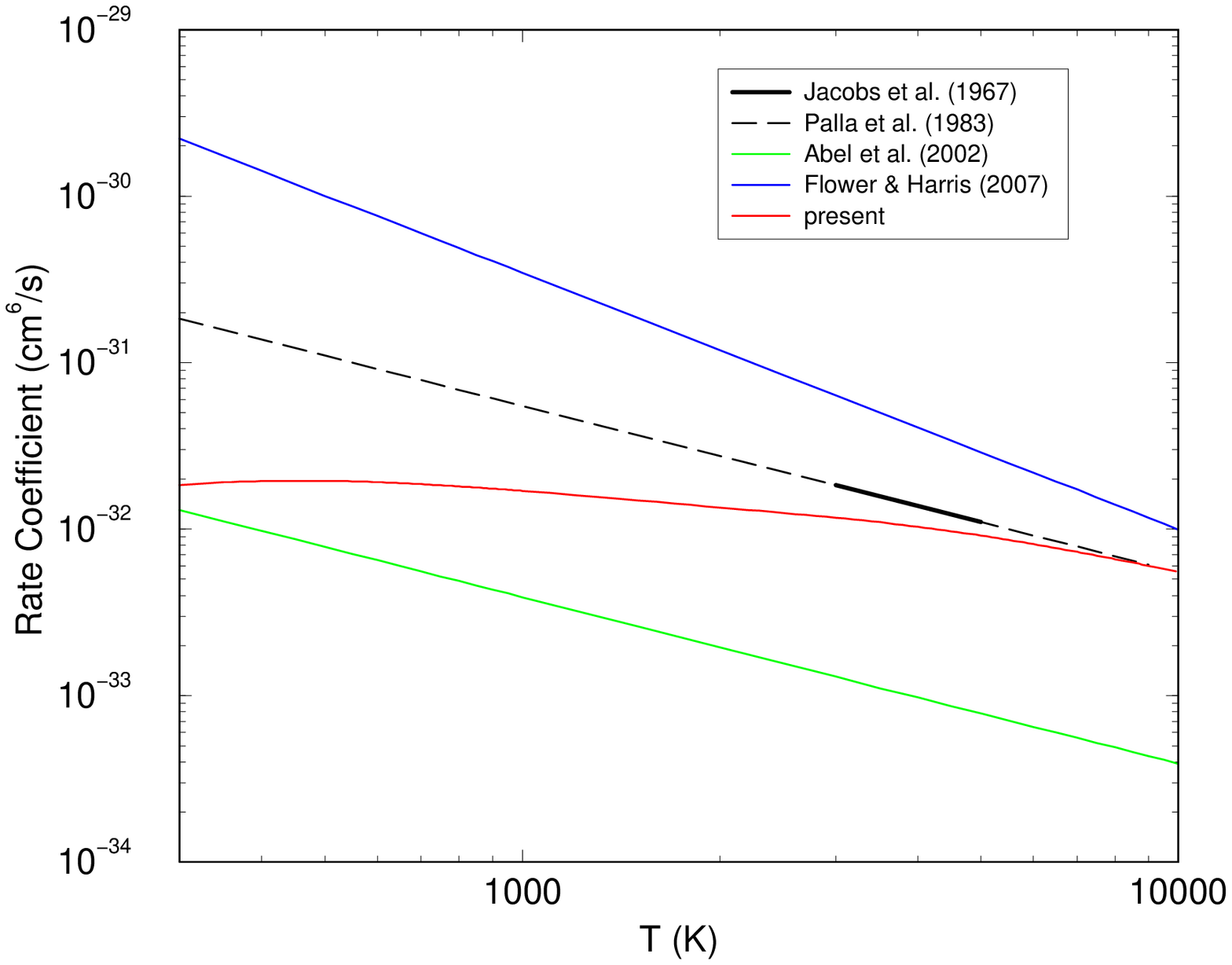,width=6in}}
\vspace*{.2in}
\footnotesize{\noindent Figure 1: TBR rate constant for H+H+H.
The present results show a much flatter temperature dependence 
than the curves of Flower and Harris \cite{flower},
Palla et al \cite{palla}, and Abel et al. \cite{abel}.
}

\end{document}